\documentclass[twocolumn,aps,prl,amssymb]{revtex4}
\usepackage{floatflt}
\usepackage{hhline}
\usepackage{graphicx}
\usepackage[figuresright]{rotating}
\usepackage{rotate}
\newcommand{\be}{\begin{equation}}
\newcommand{\ee}{\end{equation}}
\newcommand{\bea}{\begin{eqnarray}}
\newcommand{\eea}{\end{eqnarray}}

\newcommand{\aver}[1]{\langle #1 \rangle}
\begin{document}
\title{Variable size particle probing of the DLA cluster properties}

\author{A.Yu.\ Menshutin$^{1,a)}$, L.N.\ Shchur$^{1,2)}$, and
V.M.\ Vinokur$^2)$}

\affiliation{ $^{1)}$ Landau Institute for Theoretical Physics,
142432
Chernogolovka, Russia \\
$^{2)}$ Materials Science Division, Argonne National Laboratory,
Argonne, Illinois 60439, USA \\
$^a) $e-mail: \tt may@itp.ac.ru}

\begin{abstract}

We develop a technique for probing harmonic measure of the diffusion
limited aggregation (DLA) cluster surface with the variable size
particle and generate one thousand clusters with 50 million
particles using original off-lattice killing-free algorithm. Taking,
in sequence, the limit of the vanishing size of the probing
particles and then sending the growing cluster size to infinity, we
achieve the unprecedented accuracy determining the fractal
dimension, $D=1.7100(2)$ crucial to characterization of geometric
properties of the DLA clusters.
\end{abstract}

\pacs{}
\maketitle

Kinetic interfaces that evolve into two dimensional fractals via the
diverse stochastic growth processes are ubiquitous to nature.  The
systems and physical processes that exhibit kinetic roughening and
fractal structure range from firefronts and bacterial colonies to
dendrites of various nature, to domain walls in magnets and
ferroelectrics, to liquids penetrating porous media, and many
others. The study of two-dimensional fractals associated with the
kinetic roughening is an exciting and mature branch of statistical
physics -- see excellent works~\cite{HHZ,BH} for exhaustive reviews.
The generic process governing a good part of these phenomena is the
so called two dimensional aggregate growth, which is commonly
modeled as the diffusion limited aggregation (DLA)~\cite{WS} and its
generalization DBM~\cite{Pietronero} well capturing most of the
properties of the random aggregates~\cite{BH}.  Past analytical and
numerical studies advanced impressively our understanding of DLA
(see again~\cite{HHZ,BH}), yet there are still several critical
issues that remain unresolved, like the controversies related to the
multiscale and fractal nature of DLA, to name a few. One of the
central questions is the precise value of the DLA fractal dimension
$D$, the knowledge of which is crucial for full description of
geometric properties and characterization of the DLA clusters. In
particular, the exact value of $D$ is necessary for evaluating DLA
lacunarity at large growth times.

In the planar, $d{=}2$, geometry the analytical results (see for
example~\cite{Gould}) suggest that the DLA fractal dimension lies in
the range $D{=}1.67{-}1.72$; certain rational values like $D{=}5/3$,
in~\cite{Muthu}, or $D{=}17/10$, in~\cite{Hastings} were also
predicted. On the other hand, Mandelbrot~\cite{Mandelbrot} argued
that DLA clusters fill the whole space, i.e. that $D{=}2$. The
direct simulations usually produce $D{=}1.715(4)$ for the DLA
fractal dimension~\cite{tm89,ms-Os91}; the simulations~\cite{DLP}
based on the conformal mapping~\cite{HL,ms-DHOPSS} yield
$D{=}1.713(5)$.

Conventionally the fractal dimension $D$ is determined from the fit
of the growing aggregate (cluster) size, $R$, to the functional
dependence

\begin{equation}
R \propto N^{1/D}, \label{RND}
\end{equation}

\noindent where $N$ is the number of particles in the cluster.  The
cluster size, $R$, can be defined as the radius of the deposition
$R_{dep}{=}\langle r_i \rangle$, where $r_i$ is the position of the
$i$-th particle, and brackets stand for the averaging over the
ensemble of clusters.  The accuracy achieved in the determination of
$D$ in the past publications is characterized by the
fluctuations~\cite{RS,MS} ${\cal
F}_D{=}\left(\aver{D^2}{-}\aver{D}^2\right){/}\aver{D}^2$. The
latter decays as ${\cal F}_D{\propto}N^{-0.33}{\simeq}N^{-1/3}$
(rather than naively expected ${\propto}1/N\ln N$).  As a matter of
practice, this implies that in order to achieve the next level of
accuracy [one more (forth) digit] one is required to consider
clusters with the number of particles exceeding by the factor of
1000 those used in simulations of Refs.~\cite{tm89,ms-Os91,DLP}. The
clusters of such a size, about  $N{=}10^9$, are not accessible via
standard approaches due to the computational speed and computer
memory limitations.

In this Letter we develop and report on the alternative approach
allowing to reach higher accuracy of the fractal dimension
measurement. We exploit the fact that in simulations the diffusing
particles always have some finite physical size $\delta$ and that
while the true fractal dimension $D$ should not depend on $\delta$,
the results of simulations appear to depend on the size of the
particles probing harmonic measure. Our idea is to use this
dependence in order to parameterize the fractal dimensionality as a
function of $\delta$ and, upon finding $D(\delta)$ in a series of
numerical experiments, define the fractal dimension~\footnote{Our
definition of the fractal dimension does not depend upon the size of
the particles the cluster itself is composed of. Therefore, we let
the size of the particles in the cluster to be unity and measure the
size of the probing particles in the units of the cluster
particles.} as $D{=}\lim_{\delta{\to} 0}D(\delta)$. We will show
that our approach allows for a dramatic improvement in the precision
in determining $D$.

The accessibility of a particular set of the points at the cluster
interface is characterized by the probabilities $p_k$ for a particle
diffusing from the infinity to hit the cluster at this given subset,
$\Gamma_k$, of the surface. For the interface of the form of an
ideal circle (and the symmetric diffusion), the probabilities are
distributed uniformly, and the probability density is just $1/2\pi$
(the harmonic measure). In an irregular cluster most of the
probability density is concentrated around the cluster's tips,
whereas the fjords are screened (this effect is often referred to as
the ``tip effect'' and/or the ``Faraday
screening''~\cite{Pietronero,BA,BB,HGEH}). The conventional approach
to probing the harmonic measure is as follows: (i) The DLA cluster
containing $N$ particles is generated; (ii) The positions $r_i$
where the next $M$ particles hit the border of the cluster for the
first time are stored but the particles do not remain attached to
the cluster; (iii) The deposition radius is calculated as

\begin{equation}
R_{dep}(N)\approx \frac 1M \sum_{i=1}^M r_i \label{rdep},
\end{equation}

\noindent and the sum converges to the integral over harmonic
measure $R_{dep}(N){=}{\int}\!dq\, r$ for sufficiently large $M$.

\begin{figure}
\includegraphics[angle=270,width=\columnwidth]{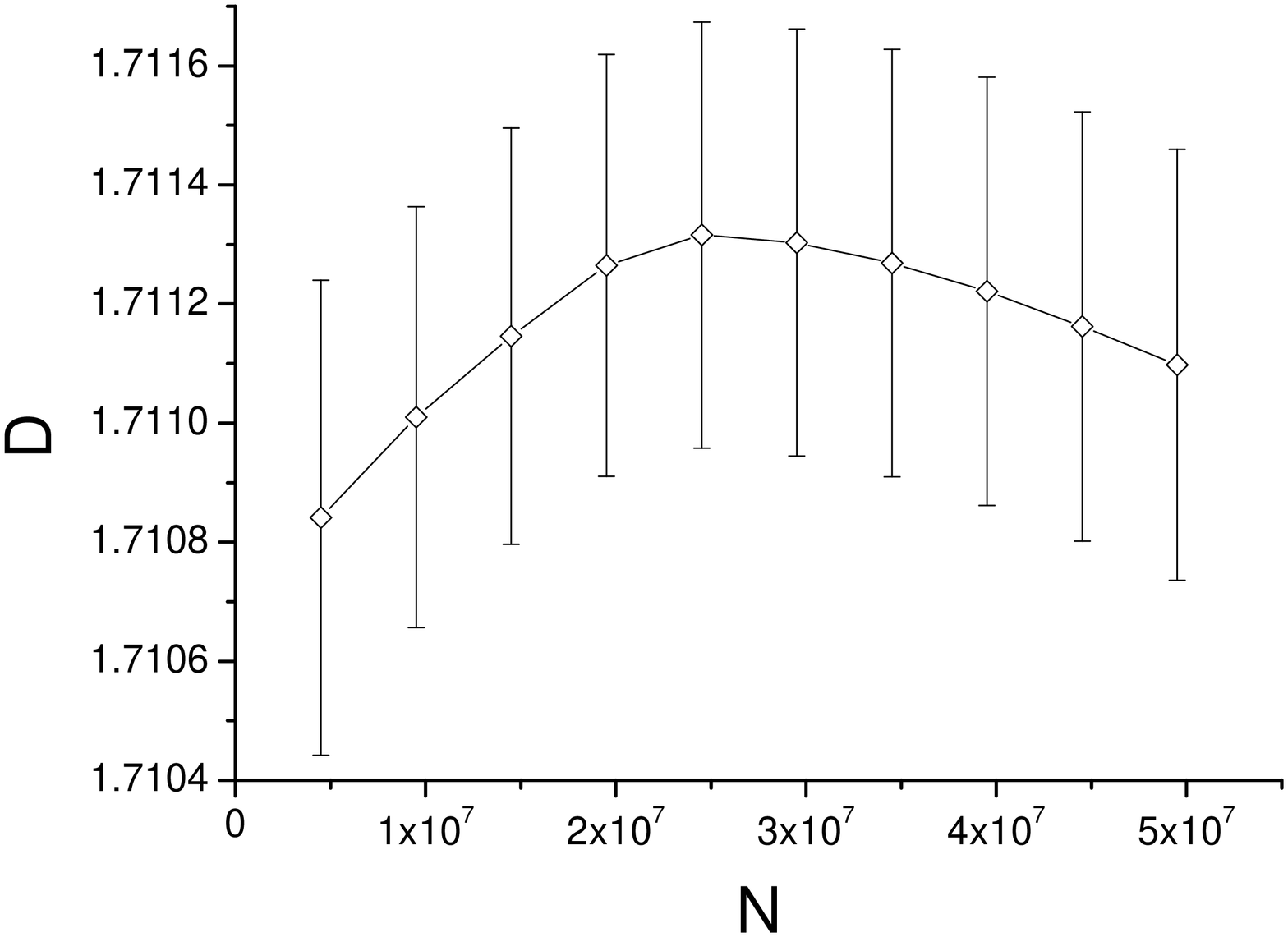}
\caption{Fractal dimension with error bars as function of the
cluster size, computed with $R_{dep}$. Solid line connecting circles
is guide for the eye.} \label{osc1}
\end{figure}

Shown in Figure~\ref{osc1} is the typical variation of $D$
calculated using relation~(\ref{RND}) with $R_{dep}$, and
representing the half period of oscillations of fractal dimension
$D$ with the size of the cluster. The oscillations are related to
the spatially nonuniform growth of the cluster: different branches
grow with the different speed, and the newborn sub-branches often
outgrow the parent branches.  The amplitude of the oscillations
decrease slowly with the cluster size thus complicating precise
determination of the value of $D$.

\begin{figure}
\includegraphics[width=\columnwidth]{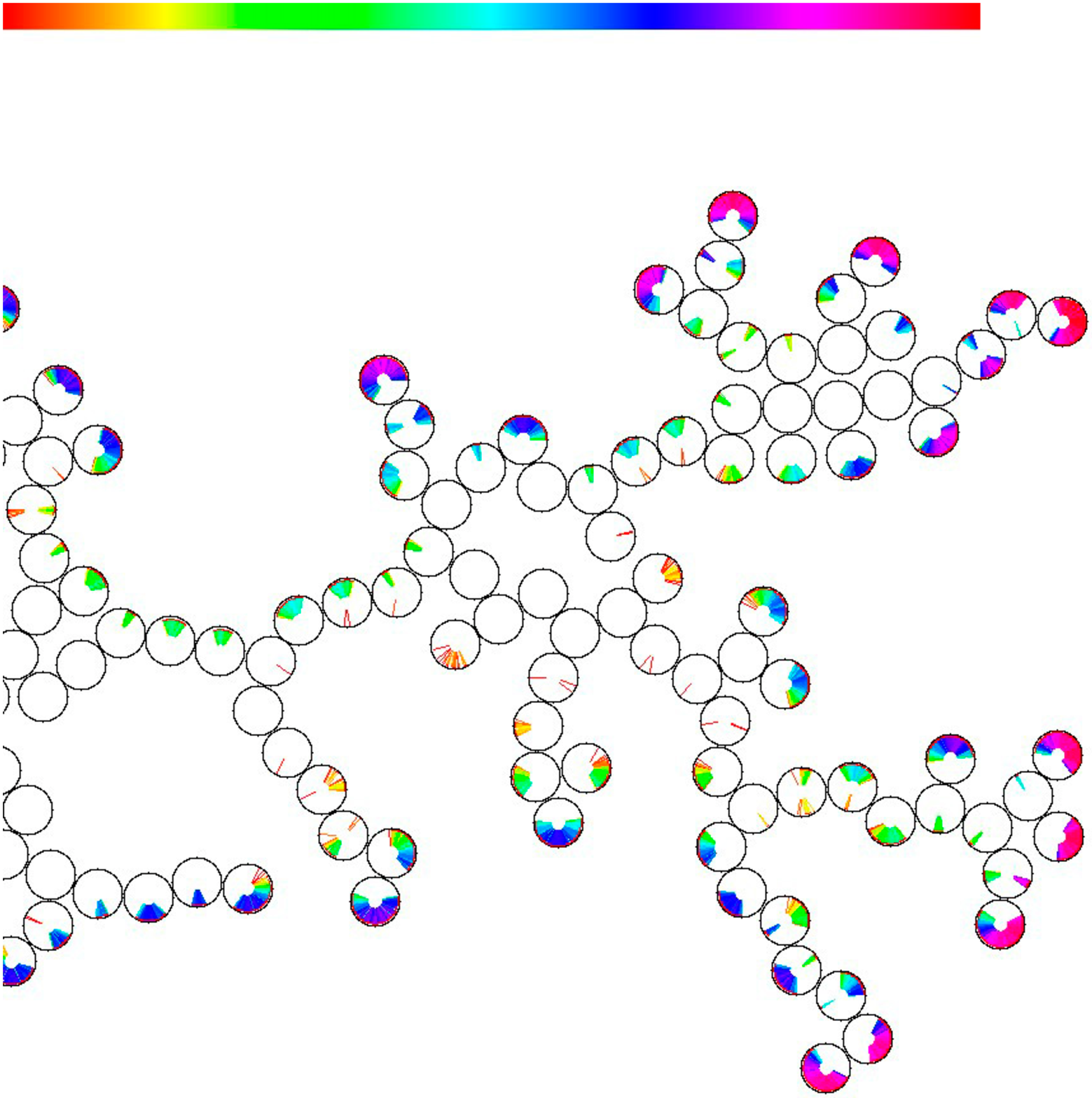}
\caption{(Color online) Fragment of the cluster with $N{=}10^4$
particles and positions of hits by probe particles with radius
$\delta{=}1$. Color is proportional to logarithm of the hit
probabilities, denoted by the color bar at the top. Probabilities
are larger near the tips and smaller inside the fjords.}
\label{pic1}
\end{figure}

Figure~\ref{pic1} shows the part of the cluster branch. The spots
where the probe particles hit the surface are marked by the color
with the intensity proportional to the logarithm of the particular
probability $p_k$ for the particle to hit the segment $\Gamma_k$.
For the sake of better visualization we choose the length of
segments equal to the one pixel of the figure.  Uncoupled colored
segments, $\Gamma_k$ (see Fig.~\ref{pic1}), constitute only the part
of the surface accessible to the diffusing particles. The total
length of the accessible cluster interface depends on the size of
the probe particles. We choose the probe particle size $\delta$ as a
parameter of the particular measurement. The value of $\delta$
controls two processes: (i) the `geometrical' process which is the
penetration of the particles inside the fjords where the geometrical
bottleneck of the fjords may prevent~\cite{BB,HGEH} sizable
particles to go through, and (ii) the probabilistic process of
screening; the latter becomes more effective with the growth of the
particle size, $\delta$.

It is important to stress the difference between the measurement of
the length of the fractal surface~\cite{fractal} and the measurement
of the harmonic measure on the surface within our approach. In the
former case, the total length of surface grows with the decreasing
scale division value~\cite{fractal} (i.e. the length of the subsets
$\Gamma_k$ are equal to the ruler size). In the latter case, the
probability $p_k$ for the particle to hit the subset $\Gamma_k$
saturates as $\delta{\to}0$.

\begin{figure}
\includegraphics[angle=270,width=\columnwidth]{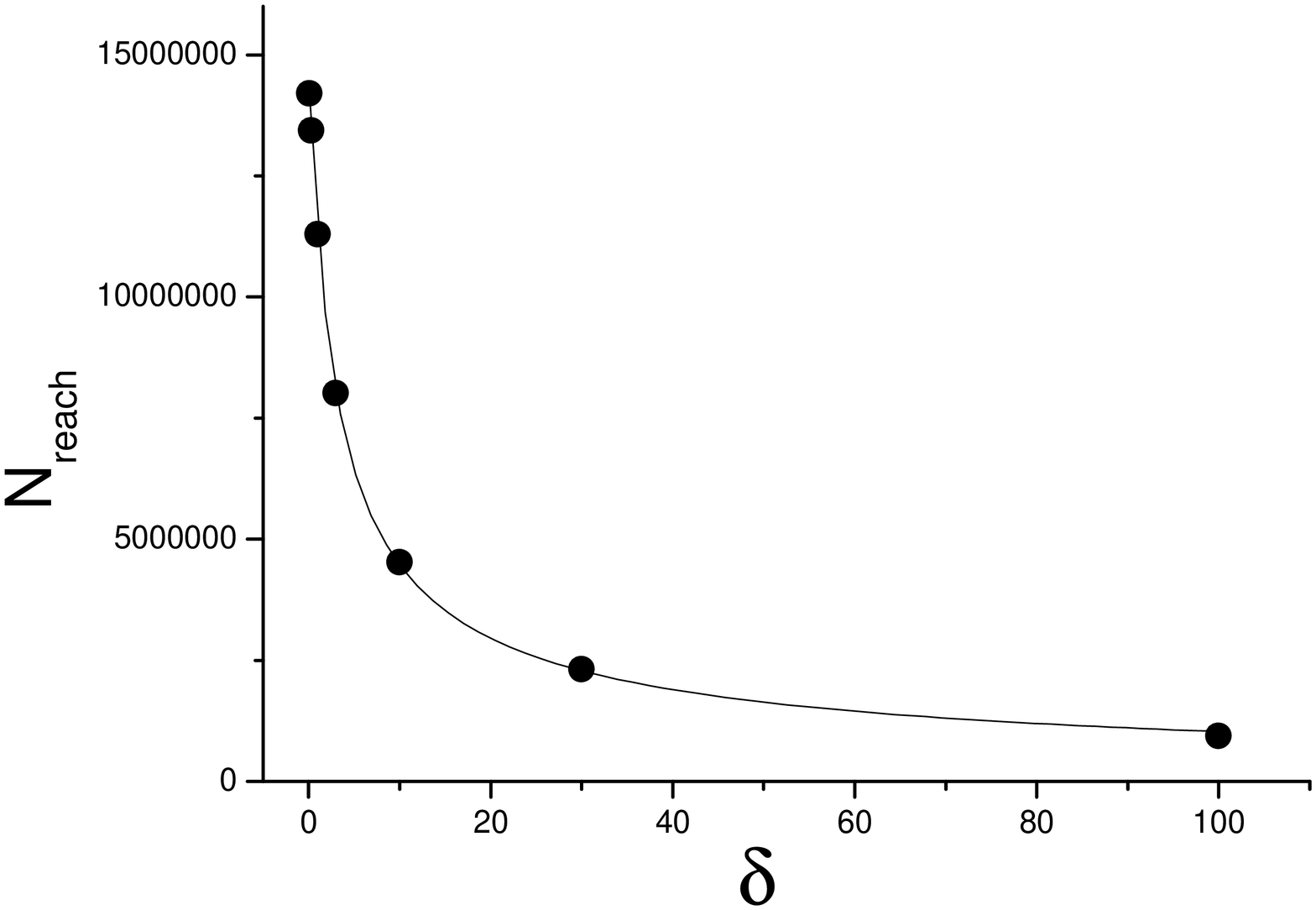}
\caption{Number of reachable sites of the cluster as function of the
probe particle radius (circles) with the fit described in the text
(solid line).}
\label{reach}
\end{figure}

Shown in Figure~\ref{reach} is the number of the surface particles
having been touched by the probe particles as function of the
probing particle size.  The solid line is given by the expression
$N_{reach}{=}N_{surf}/(1{+}\delta/\delta_0)^\alpha$, with
$\alpha{=}0.69(2)$ and $\delta_0{=}2.2(1)$.  The limit of vanishing
size $\delta{\to}0$ gives the limiting number of the reached
particles as the total number of surface particles,
$N_{surf}{\approx}14,662,525$ (out of total $20$ millions in the
cluster).

\begin{figure}
\includegraphics[angle=0,width=\columnwidth]{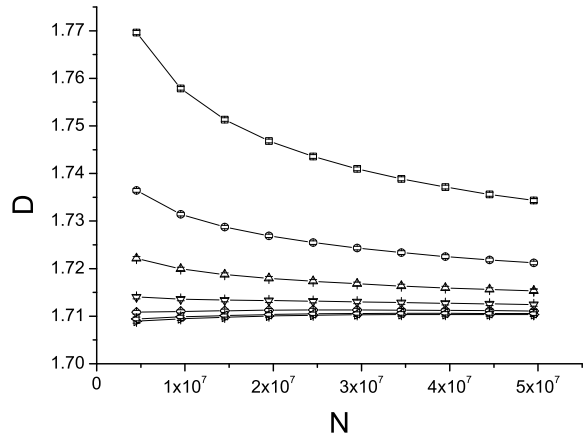}
\caption{Fractal dimension (symbols) as function of $N$ for
$\delta=$0.1, 0.3, 1, 3, 10, 30, 100 from bottom to top. Solid lines
connecting circles is guide for the eye.} \label{dnd}
\end{figure}

\begin{figure}
\includegraphics[angle=0,width=\columnwidth]{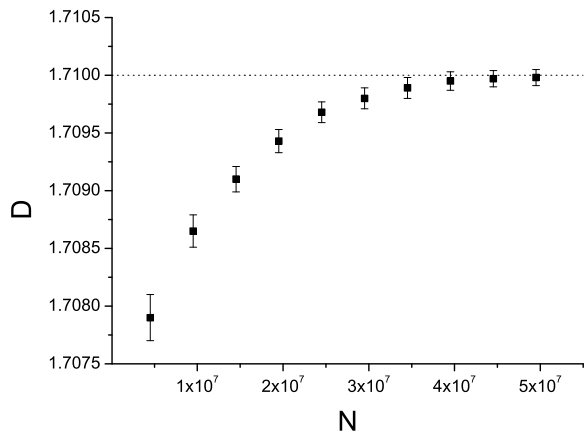}
\caption{Fractal dimension as function of $N$ for the limiting value
of $\delta=0$. It reach value $D=1.7100(2)$ (plotted with the dotted
line) in the limit of large cluster size $N$.} \label{dn}
\end{figure}

The dependence of $D$ upon $N$ for different $\delta$ is shown in
Fig.~\ref{dnd}.  As we have already mentioned, the dependence on $N$
within the given interval of $N$ and for the fixed $\delta$ is not
monotonic for $\delta{<}3$. Thus, we fit results by the formula

\begin{equation}
D(\delta;N)=D(N)+A\delta^\beta \label{expr-dn}
\end{equation}

\noindent and take the limit of $\delta{\to}0$ for the fixed values
of the cluster size $N$. The resulting values of $D(N)$ are shown in
Fig.~\ref{dn}. Note that now the dependence $D(N)$ has become
monotonic.  This can be understood as the result of the effective
averaging the competition between the growing branches out.  We have
examined the ensemble of 1000 clusters with 50 million particles for
each value of $\delta$. The number of particles $M$ in each event of
the measurement necessary to achieve the  desired accuracy in
$R_{dep}$ of about 0.1\%, was typically several tens of thousands.
We have found no difference in results when using larger number of
probe particles. The error bars represent combined errors from both
the ensemble average and the fit. Thus from Fig.~\ref{dn} one finds
at the end of the day the ultimate value of the fractal dimension,
$D{=}1.7100(2)$. This is an unprecedented accuracy in the
measurements of $D$ exceeding by the order of magnitude the results
known from the literature.

It is instructive to use the same ensemble of clusters in order to
evaluate the fractal dimension via the traditional approach
introducing the deposition radius as $R_{dep}{=}\langle r \rangle$,
the mean square displacement $R_2{=}\sqrt{\langle r^2\rangle}$, and
the radius of gyration
$R_{g}{=}\sqrt{\frac{1}{N}\sum_{i{=}1}^N\langle r^2\rangle}$. The
results of the fit to the form~(\ref{RND}), where $R$ stands for one
of the above radii, are presented in Table~\ref{tab1}. Note, that
all the obtained values of $D$ are larger than that of
$D{=}1.7100(2)$ derived by the method of present work.

Since DLA clusters grow randomly, a center of the cluster mass
$R_{M}{=}\sqrt{\frac{1}{N}\left|\sum_{i{=}1}^N r_i\right|}$ performs
random walks in the plane. Accordingly, the distance from the
original position of a seed particle to the center of mass grows as
$\sqrt{N}$ and the number of particles is proportional to the time
of the random walk~\footnote{There are some attempts in literature
of extracting the fractal dimension of the cluster using fit
$R_M{\propto} N^{1/D}$.  However no justification for such fit is
available to the best of our knowledge. One rather expects the
center of cluster mass to perform a conventional random walk.}. The
average value of $R_M$ for $N{=}5{\cdot}10^7$ is $\langle
R_M\rangle{\approx}1000$ which should be compared to $\langle
R_g\rangle{\approx}30000$ and the penetration depth
$\xi{\approx}7000$. The average angular position of the center of
mass (averaged over an ensemble) is also a stochastic function of
time, although at each realization of the cluster angular
correlations are observed (some of the branches grow faster within a
given time interval~\cite{MSV-long}).

The important question now is whether the choice of the coordinate
frame influences the final result and the value of the fractal
dimensionality.  Indeed, when determining $D$ one can choose the
origin at either (i) the position of the seed particle, or (ii) at
the (evolving with time) position of the center of gravity, or else
(iii) at the 'center of the charge gravity' (the latter is most
appropriate for the DMB case).  We have performed averaging
according to Eq.~(\ref{rdep}) finding $D$ from the fit to $\tilde
R_{dep}$, $\tilde R_2$, and $\tilde R_g$, placing the origin of the
reference frame to the seed particle (left column of the
Table~\ref{tab1}) and to the center of gravity (right column).  One
sees that for all the quantities the choice of the reference frame
is irrelevant (within the accuracy of the computation), although the
ensemble of clusters should be large enough to insure this
convergence.

\begin{table}
\begin{tabular}{|l|l|l|} \hline
 & $r$ to seed & $r$ to center-of-mass   \\ \hline
$R_{dep}$ & 1.7098(12) & 1.7111(6) \\
$R_2$   &   1.71155(56) & 1.71149(54) \\
$R_g$   &   1.71149(30) & 1.71133(30) \\ \hline
\end{tabular}
\caption{Fractal dimension estimated by the standard method using
the reference frames centered at the seed particle (left column) and
at the center of gravity (right column) respectively. } \label{tab1}
\end{table}

In conclusion, we have developed a technique for the high precision
analysis of the geometrical properties of DLA clusters, in
particular the evaluation of its lacunarity in the long time limit.
Our approach offers a perfect tool for further advance in our
understanding of the fjord screening behavior. In particular, the
long standing problems of the behavior of the $p_{min}$, which is
the minimal growth probability (in DBM language) or probability to
hit given segment of the cluster surface (in DLA language) can be
efficiently addressed via the developed approach. The behavior of
the latter probability is related to the phase transition in the
multifractal spectrum~\cite{LS}. To the best of our knowledge, this
probability was estimated only within the ``tunnel
configuration''~\cite{BA} technique and had never been measured in
the simulations of the ``typical configuration''~\cite{LAS}.

We are pleased to thank S. Korshunov for important and enlightening
discussions. This work is supported by the US DOE Office of Science
under contract No. W31-109-ENG-38 and by the Russian Foundation for
Basic Research.


\begin{thebibliography}{99}

\frenchspacing

\bibitem{HHZ} T. Halpin-Healy and Y.C. Zhang, Physics Reports {\bf 254}, 215 (1995)

\bibitem{BH} A. Bunde and S. Havlin, eds.,
  {\it Fractals and Disordered Systems}
(Springer, Berlin, 1996).

\bibitem{WS} T.A. Witten and L.M. Sander, Phys. Rev. Lett. {\bf 47}, 1400 (1981).

\bibitem{Pietronero} L. Niemeyer, L. Pietronero, H.J. Wiesmann,
Phys. Rev. Lett., {\bf 52}, 1033 (1984).

\bibitem{Gould} H. Gould, F. Family, and H.E. Stanley, Phys. Rev. Lett. {\bf 50},
686 (1983).

\bibitem{Muthu} M. Muthukumar, Phys. Rev. Lett. {\bf 50}, 839
(1983).

\bibitem{Hastings} M.B. Hastings, Phys. Rev. E {\bf 55}, 135 (1997).

\bibitem{Mandelbrot} B. Mandelbrot, Physica A {\bf 191}, 95 (1992).

\bibitem{tm89} S. Tolman and P. Meakin, Physica A {\bf 158}, 801 (1989);
Phys. Rev. A {\bf 40}, 428 (1989).

\bibitem{ms-Os91} P. Ossadnik, Physica A {\bf 176} (1991) 454, ibid.
{\bf 195}, 319 (1993).

\bibitem{DLP} B. Davidovitch, A. Levermann, and I. Procaccia,
Phys. Rev. E {\bf 62}, R5919 (2000).

\bibitem{HL} M.B. Hastings and L.S. Levitov, Physica D {\bf 116}, 244 (1998).

\bibitem{ms-DHOPSS} B. Davidovitch, H.G.E. Hentschel, Z. Olami, I.
Procaccia, L.M. Sander, and E. Somfai, Phys. Rev. E {\bf 59}, 1368
(1999).

\bibitem{RS} T.A. Rostunov, L.N. Shchur, JETP 95, 145 (2002)

\bibitem{MS}A. Yu. Menshutin, L. N. Shchur, Phys. Rev. E {\bf 73}, 011407
(2006).

\bibitem{BA} R. Blumenfeld, A. Aharony, Phys. Rev. Lett. {\bf 62},
2977 (1989)

\bibitem{BB} R.C. Ball, R. Blumenfeld, Phys. Rev. A {\bf 44}, R828
(1991).

\bibitem{HGEH} H.G.E. Hentschel, Phys. Rev. A {\bf 46}, R7379
(1992).

\bibitem{fractal} B.B. Mandelbrot, Science {\bf 155}, 636 (1967).

\bibitem{MSV-long} A.Yu. Menshutin, L.N. Shchur, and V.M. Vinokur,
unpublished.

\bibitem{LS} J. Lee, H.E. Stanley, Phys. Rev. Lett. {\bf 61},
2945 (1988)

\bibitem{LAS} J. Lee, P. Alstr\"om, and H.E. Stanley, Phys. Rev. Lett. {\bf 62},
3013 (1989)

\end{thebibliography}
\end{document}